\documentstyle[12pt]{article}

\def\hybrid{\topmargin -20pt    \oddsidemargin 0pt  
        \headheight 0pt \headsep 0pt  
        \textwidth 6.25in       
        \textheight 9.5in       
        \marginparwidth .875in  
        \parskip 5pt plus 1pt   \jot = 1.5ex}  

\def\noi{\noindent}  
  
\def\baselinestretch{1.2}  
  
\catcode`\@=11  
  
\def\marginnote#1{}  
\def\draftlabel#1{{\@bsphack\if@filesw {\let\thepage\relax  
   \xdef\@gtempa{\write\@auxout{\string  
      \newlabel{#1}{{\@currentlabel}{\thepage}}}}}\@gtempa  
   \if@nobreak \ifvmode\nobreak\fi\fi\fi\@esphack}  
        \gdef\@eqnlabel{#1}}  
\def\@eqnlabel{}  
\def\@vacuum{}  
\def\draftmarginnote#1{\marginpar{\raggedright\scriptsize\tt#1}}  
  
\def\draft{\oddsidemargin -.2truein  
        \def\@oddfoot{\sl preliminary draft \hfil  
        \rm\thepage\hfil\sl\today\quad\militarytime}  
        \let\@evenfoot\@oddfoot \overfullrule 3pt  
        \let\label=\draftlabel  
        \let\marginnote=\draftmarginnote  
   \def\@eqnnum{(\theequation)\rlap{\kern\marginparsep\tt\@eqnlabel}%
\global\let\@eqnlabel\@vacuum}  }  
  
\def\preprint{\twocolumn\sloppy\flushbottom\parindent 2em  
        \leftmargini 2em\leftmarginv .5em\leftmarginvi .5em  
        \oddsidemargin -.5in    \evensidemargin -.5in  
        \columnsep .4in \footheight 0pt  
        \textwidth 10.in        \topmargin  -.4in  
        \headheight 12pt \topskip .4in  
        \textheight 6.9in \footskip 0pt  
        \def\@oddhead{\thepage\hfil\addtocounter{page}{1}\thepage}  
        \let\@evenhead\@oddhead \def\@oddfoot{} \def\@evenfoot{} }  
  
\def\numberbysection{\@addtoreset{equation}{section}  
        \def\theequation{\thesection.\arabic{equation}}}  
  
\def\underline#1{\relax\ifmmode\@@underline#1\else  
        $\@@underline{\hbox{#1}}$\relax\fi}

\def\titlepage{\@restonecolfalse\if@twocolumn\@restonecoltrue  
\onecolumn  
     \else \newpage \fi \thispagestyle{empty}\c@page\z@  
     \def\thefootnote{\fnsymbol{footnote}} }

\def\endtitlepage{\if@restonecol\twocolumn \else \newpage \fi  
        \def\thefootnote{\arabic{footnote}}  
        \setcounter{footnote}{0}}  
  
\catcode`@=12  
\relax  
  
\def\figcap{\section*{Figure Captions\markboth  
        {FIGURECAPTIONS}{FIGURECAPTIONS}}\list  
        {Figure \arabic{enumi}:\hfill}{\settowidth\labelwidth{Figure  
999:}  
        \leftmargin\labelwidth  
        \advance\leftmargin\labelsep\usecounter{enumi}}}  
 \relax  
\def\tablecap{\section*{Table Captions\markboth  
        {TABLECAPTIONS}{TABLECAPTIONS}}\list  
        {Table \arabic{enumi}:\hfill}{\settowidth\labelwidth{Table  
999:}  
        \leftmargin\labelwidth  
        \advance\leftmargin\labelsep\usecounter{enumi}}}  
 \relax  
\def\reflist{\section*{References\markboth  
        {REFLIST}{REFLIST}}\list  
        {[\arabic{enumi}]\hfill}{\settowidth\labelwidth{[999]}  
        \leftmargin\labelwidth  
        \advance\leftmargin\labelsep\usecounter{enumi}}}  
 \relax  
\makeatletter  
\newcounter{pubctr}  
\def\publist{\@ifnextchar[{\@publist}{\@@publist}}  
\def\@publist[#1]{\list  
        {[\arabic{pubctr}]\hfill}{\settowidth\labelwidth{[999]}  
        \leftmargin\labelwidth  
        \advance\leftmargin\labelsep  
        \@nmbrlisttrue\def\@listctr{pubctr}  
        \setcounter{pubctr}{#1}\addtocounter{pubctr}{-1}}}  
\def\@@publist{\list  
        {[\arabic{pubctr}]\hfill}{\settowidth\labelwidth{[999]}  
        \leftmargin\labelwidth  
        \advance\leftmargin\labelsep  
        \@nmbrlisttrue\def\@listctr{pubctr}}}  
 \relax  
\makeatother  
\newskip\humongous \humongous=0pt plus 1000pt minus 1000pt

\newif\ifdtup

\font\Scbig=cmss10 scaled\magstep1  
\font\Scscr=cmss8 scaled\magstep1  
\font\Scscrscr=cmss8  
\newfam\Scfam  
\textfont\Scfam=\Scbig  
\scriptfont\Scfam=\Scscr  
\scriptscriptfont\Scfam=\Scscrscr

\relax  
\hybrid  
\def\lvm{\leavevmode\hbox to\parindent{\hfill}}  

\def\thefootnote{\arabic{footnote}}
\def\BE{\begin{equation}}  
\def\EE{\end{equation}}  
\def\BA{\begin{eqnarray}}  
\def\EA{\end{eqnarray}}

\def\tt{\bar\tau}  
  
\def\lvm{\leavevmode\hbox to\parindent{\hfill}}  
\def\bar{\overline}

\def\BE{\begin{equation}}  
\def\EE{\end{equation} \vskip 0.30\baselineskip}  
\def\BA{\begin{array}}  
\def\EA{\end{array}}  
  
\def\noi{\noindent}  
  
\def\frac#1#2{{\textstyle{{#1}\over{#2}}}}

\newif\ifold \oldtrue   
\let\ssection=\section  
\def\section{\setcounter{equation}{0}\ssection}  
  
\begin{document}  
\renewcommand{\theequation}{\arabic{equation}}  
\newcommand{\beq}{\begin{equation}}  
\newcommand{\eeq}[1]{\label{#1}\end{equation}}  
\newcommand{\ber}{\begin{eqnarray}}  
\newcommand{\eer}[1]{\label{#1}\end{eqnarray}}  
\begin{titlepage}  
\begin{center}  
   
\hfill {\ }{\ }{\ }{\ }\\
\hfill physics/0512062
\vskip 2in
   
{\large \bf A Solution to the Fermi Paradox: The Solar System,
Part of a Galactic Hypercivilization?} 
\vskip 1.5in 
 
{\large \bf Beatriz Gato-Rivera} \\

\vskip 1in

  

\vskip 1.8in  
  
{\large \bf World Mystery Forum 2005, Interlaken (Switzerland)}   

\end{center}
\vskip .4in
{\large \bf {\ \ \ }November 2005}

\end{titlepage}  
  
\def\baselinestretch{1.2}  
\baselineskip 17 pt  

\section{Introduction}\lvm

This talk is about the possibility that
the Solar System belongs to the territory of a hypercivilization
spanning our galaxy or a large region of it. I will start introducing the
Fermi Paradox (why we do not see aliens around?) and some of its 
solutions. Then I will present my own solution which includes two 
proposals called the Subanthropic Principle and the Undetectability 
Conjecture. This solution states that, at present, all typical galaxies
like ours are already colonized by very advanced technological 
civilizations spread through large regions or the whole galaxies, many
of them containing primitive subcivilizations like ours. After discussing some 
consequences of this solution for our planet and our civilization I will make 
some comments on recent, very popular theories in the scientific community of 
Particle Physics and Cosmology. These theories, known as `brane worlds', 
assume that our visible Universe with three space dimensions is 
embedded in a much larger Cosmos with more space dimensions. 
Therefore it would be most natural if other universes would also exist 
located along the extra space dimensions. As a result, these 
theories open up enormous possibilities regarding the visitation or 
colonization of the Solar System by alien civilizations, strengthening 
the Fermi Paradox. Finally, in the appendix I have included some questions 
and answers that came up during this Forum.

\section{The Fermi Paradox}\lvm  
  
In the summer 1950, in Los Alamos, the nuclear physicists 
Enrico Fermi, Edward Teller and other colleagues brought up
the subject of unidentified flying objects (UFO's) while
having lunch. This topic was very popular at that time. 
After a while, when they had changed subjects Fermi suddenly asked:
Where is everybody? Performing fast mental computations, Fermi had reached the
conclusion that alien civilizations should have been around visiting Earth for many
thousands or millions of years. Therefore, why we do not see them?
This is the Fermi Paradox. 

Although Fermi never 
explained how he made his computations, nor gave an
estimate of the number of civilizations which should have visited Earth, he had to
rely on arguments like these: In our galaxy there are thousands of millions of 
stars much older than the Sun, many of them thousands of millions of years older 
(in the `habitable zone' of the galaxy they are on average one thousand 
million years older \cite{LFG}). 
Therefore many civilizations must have arisen in our galaxy before ours and 
a fraction of them must have expanded through vast regions 
or even through the whole galaxy.

Some other arguments pointing in the same direction involve estimates 
about the lifetime of the second generation stars, inside of which the chemical 
elements of organic matter are made, and also involve estimates of
the total time necessary for a technological civilization to colonize, or visit, the
whole galaxy. Regarding the second generation stars, they are formed only two
million years after the supermassive first generation stars (these burn out 
exploding as supernovae in one million years only and  it takes 
another million years for the debris to form new stars). Therefore the appearance 
of organic matter in our galaxy could have happened several thousands of 
millions of years before the Sun came into existence. As to the total time 
necessary to colonize, or visit, the whole galaxy by a technological
civilization, conservative computations of diffusion modeling give estimates from 
5 to 50 million years \cite{Sci}, which is a cosmologically short 
timescale\footnote{ During this Forum Claudio Maccone, member of the 
International Academy of Astronautics, has presented a mathematical model
that predicts around 150 million years, which is still a short time.}. Besides 
these considerations, the fact that life on Earth started very early supports the 
views, held by many scientists, that life should be abundant in the Universe.

\section{Solutions to the Fermi Paradox}\lvm

Many solutions have been proposed to the Fermi Paradox. I classify them
as expansionist and non-expansionist. The non-expansionist solutions are based 
on the assumption that technological civilizations do not expand beyond a small 
neighborhood of the galaxy. The most popular of these solutions are the 
following ones:

\begin{itemize}

\item
Interstellar travel is not possible no matter the scientific and technological
level reached by a civilization. 

\item 
Generically, advanced civilizations have little or no interest in expanding 
through large regions of the galaxy.

\item 
Technological civilizations annihilate themselves, or disappear by natural 
catastrophes, before having the chance to spread through large regions of the galaxy.

\end{itemize}

\vskip .2in

On the other hand, the most popular expansionist solutions to the Fermi Paradox,
based on the assumption that generically technological civilizations 
do expand through large regions of the 
galaxy, make use of one or more of the following arguments:

\begin{itemize}

\item 
Alien civilizations do visit Earth at present times, for different purposes, and/or 
have visited Earth in the past. In this respect it is 
remarkable the fact that Francis Crick, one of the discoverers of the DNA 
structure, proposed in the mid-seventies that life on Earth could have been
inseminated on purpose by alien intelligences\footnote{It is less known that
several years before Crick, in 1960, the astronomer Thomas Gold suggested,
during a congress in Los Angeles, that space travellers could have brought
life to Earth some thousands of millions of years ago. Curiously, this claim
was also made, with a detailed description of `the facts',  
by the `popular metaphysics' writer T. Lobsang Rampa in the book 
The Hermit (1971).}. Besides, some scientists as well as countless authors 
of popular books, have speculated that some UFO's could be true alien 
spacecrafts whereas some `gods descending from the sky', abundant
in many ancient traditions, could have been just alien astronauts
(see for example \cite{PBD} \cite{Deardorff}).

\item 
Advanced alien civilizations might have strong ethical codes against interfering
with primitive life-forms \cite{Sagan}.

\item 
Advanced aliens ignore us because of lack of interest due to our low primitive 
level. For example Robert Jastrow, ex-director of Mt. Wilson Observatory, 
claims \cite{RJ} that, on average, advanced civilizations should consider us 
as larvae due to the fact that they should be thousands of millions of years
ahead of us.... and who would be interested in communicating with larvae?

\item 
Alien civilizations have not reached us yet because intelligent life is extremely 
difficult to emerge. Otherwise alien civilizations would necessarily be here.
As a result we could find ourselves among the most evolved
technological civilizations in our galaxy or we could even be the only one.

\end{itemize}

\vskip .2in 

Besides these simple solutions there are many more exotic proposals. 
For example, a rather drastic expansionist solution is given by the 
theoretical physicist Cumrun Vafa, at Harvard University,
who thinks that the fact that we do not see aliens around could be the first proof 
of the existence of brane worlds: all advanced aliens would have emigrated to 
better parallel universes \cite{CuVa}. 

\section{My Solution to the Fermi Paradox}\lvm

Two years ago I made a proposal for solving the Fermi Paradox \cite{article}:
{\it At present, all the typical galaxies of the Universe are already
colonized (or large regions of them) by advanced civilizations. In the vast 
territory of these hypercivilizations a small proportion of their individuals
belong to primitive subcivilizations, like ours}. That is, I put forward
the possibility that our small terrestrial civilization is embedded in a large
hypercivilization unknowingly and this situation should be common in all typical
galaxies. 

The primitive subcivilizations would know or ignore their low 
status depending, most likely, on the ethical standards  
of the advanced civilization in which they are immersed. If the standards 
are low, the individuals of the primitive subcivilizations will be surely abused 
in many ways and therefore will be painfully aware of their low status. If the 
ethical standards of the advanced individuals are high instead, 
then they will respect the natural evolution (social, cultural) 
of the primitive subcivilizations, treating them `ecologically' as some
kind of protected species. In this case, which could well describe the situation 
of the terrestrial civilization, most primitive individuals would be completely 
unaware of the existence of the large hypercivilization.

\vskip .2in 

Now some remarks follow:

\begin{itemize}

\item 
Needless to say, the primitive individuals would not be considered as citizens
of the hypercivilization and direct open contact would completely destroy 
the primitive subcivilization. 

\item 
The `alien visitors', from the viewpoint of the primitive individuals,
would not be so from the viewpoint of the advanced individuals from 
the hypercivilization because 
they rather would be visiting, or working in, their own territory. (The
advanced civilizations would surely have underground and/or submarine
bases in their primitive planets for military and scientific purposes.)

\item 
I distinguish between aggressive and non-aggresive advanced civilizations.
I do not believe that advanced civilizations must have ethical codes
concerning primitive life-forms.

\item 
The fact that our civilization has never been attacked by aggressive aliens, 
as far as history knows, could well be a clue that we belong to a non-aggressive 
advanced civilization which protects planet Earth, as part of its territory.

\end{itemize}

\vskip .2in 

If this scenario is true for our civilization, then the {\it Subanthropic 
Principle} \cite{article} would also hold:
{\it We are not typical among the intelligent observers from the Universe.
Typical civilizations of typical galaxies would  be hundreds of thousands, 
or millions, of years more evolved than ours and, consequently, typical 
intelligent observers would be orders of magnitude more intelligent than 
us}\footnote{This means that, if we `score' 10, then they would score
100 or 1.000 or 10.000, etc.}.  

\vskip .2in 
One may argue against this principle saying that a much higher level of 
science and technology does not necessarily imply a much  higher
intelligence or brain capacities. My answer to this objection is that there
are two main reasons to think in this way. The first is simply natural biological
evolution as we see it in our planet. The individuals of our species
(homo sapiens sapiens) are more intelligent than our ancestors and these
were more intelligent than their own ancestors, etc. Since there is no
reason why this process should stop with us, it is fully realistic to 
expect that individuals of much older civilizations must be genetically
more intelligent than us. But there is an even stronger argument supporting
the views that the older the civilization, the more intelligent the
individuals: As the civilizations would reach some 
mastery in the field of genetic engineering, the general tendency would be to
`improve' themselves, that is their own species (among many others species) 
giving rise to an acceleration of biological evolution at unimaginable rates. 

\vskip .2in 
Now there is a crucial question. If the Solar System is part of a large
hypercivilization, why we do not detect any signs of intelligence from
outer space at all? My answer is related to my views that aggressive 
advanced civilizations must exist: {\it Generically, all advanced enough
civilizations camouflage their planets for security reasons, because of the 
existence of aggressive advanced civilizations, so that no sign of civilization 
(or any other form of life) can be detected by external observers, who  
would only obtain distorted data for disuasion purposes}.  This hypothesis
I call the {\it Undetectability Conjecture} \cite{article}.

 \vskip .2in  
Observe that this conjecture predicts a rather low 
probability of success for the SETI (search for extraterrestrial intelligence)
project. The reason is that only {\it primitive} civilizations able to produce
electromagnetic emissions would be susceptible to being detected by the SETI
antennas. But after reaching the level of producing electromagnetic emissions
it would take only a few hundred years for a civilization to learn to hide
themselves from external observers, becoming undetectable. 
As a result, the period of detectability of an average civilization could be
very short and the probability that a primitive civilization, like ours, 
detects another primitive civilization would be negligible. For example, it
could have happened that planet Earth received the last TV or radio programs
from another planet 200.000 years ago, for a period of about 500 years. 

Observe also that, if  the Undetectability Conjecture turns out to be true, 
then we cannot be sure whether the terrestrial civilization is the unique 
civilization inhabitating the Solar System, as we firmly believe. The 
reason is that in the astronomical observations of planets and satellites 
we scientists assume that there are no intelligent beings there 
manipulating the data that we receive, and then we conclude that there is 
no signal of intelligent life, {\it as the data prove}. But this assumption could
turn out to be wrong because advanced civilizations would be 
technologically able to fool our telescopes, detectors and space probes, 
and would not allow themselves to be detected.

\vskip .2in  
The relevance of this solution to the Fermi Paradox for our planet and, 
especially, for our civilization, depends strongly on when, how long ago,
the Solar System was taken over by extraterrestrial hypercivilizations
(this could have happened more than once since our galaxy is very old).

\begin{itemize}

\item 
If the Solar System was visited already thousands of millions of years ago:  
Planet Earth could have undergone insemination procedures, together with 
many other `promising' planets, as suggested by Thomas Gold and Francis 
Crick. As a result, all terrestrial living beings would have common building 
blocks of DNA with the living beings of thousands of other planets which would
have undergone similar insemination procedures with the same bacteria.
Consequently, many plants and animals, including our own species, could 
have been brought to Earth (at any time in history or prehistory) and their 
extraterrestrial origin would be impossible to detect by any biologist or 
geneticist. Moreover, at present, we probably would have countless
`relatives' in many planets, what has advantages (possibility of positive, 
closer relationships) but also disadvantages (predators). 
Not to mention the high probability that many past and present species, 
including ourselves, could have resulted from genetic manipulation
simply as a routine procedure for the improvement of the planet. (I am
convinced that advanced and not-so-advanced civilizations widely make
use of genetic engineering to improve all the species around, including
themselves.)

\item 
If the Solar System was first encountered only when life on Earth was well 
developed (for example, a few million years ago): It is still very 
probable that many present species would have resulted from genetic 
manipulation by the `owners' of the Solar System for the general 
improvement of the planet; that is, among many other improvements like 
getting rid of unwanted species or ameliorations of geological type.
(Observe that the sudden appearance of the homo sapiens sapiens species 
some 40.000 years ago is still a complete mystery for science.)

\item 
If the Solar System was first encountered only after homo 
sapiens sapiens had appeared: This hypercivilization 
must have found our species `good enough' so that they did not try any 
genetic improvements on us. This does not mean, however, 
that they would have adopted a complete non-interference policy with 
respect to human affairs.

\end{itemize}
\vskip .2in 

In fact, if my scenario is correct and the Solar System belongs to the 
territory of a non-aggressive hypercivilization, then my guess is that 
this civilization treats us as a protected species and cares about us. 
As a result, they could well have decided to help us, discretely, many 
times in history and prehistory regarding our social, cultural, 
scientific, etc. evolution, apart from protecting us and the planet from 
several dangers of various types: large meteorites, predators from
outer space, nuclear accidents, extremely devastating earthquakes, etc. 

Therefore my solution to the Fermi Paradox is compatible with the 
speculations that some UFO's could be true alien spacecrafts, whereas 
some `gods descending from the sky', for the benefit of mankind, 
could have been alien astronauts. However, I would not call them 
alien astronauts but extraterrestrial scientists and extraterrestrial
military personnel instead, since working in their own territory, they 
would not be truly astronauts nor aliens. 

My solution is also compatible with the possibility of more general close 
contacts between individuals of the hypercivilizations and primitive 
individuals like us, besides the `gods descending from the sky' issue.
I have identified three major causes or reasons which could motivate 
individuals of advanced civilizations to seek interactions or relationships 
with primitive individuals: scientific purposes, entertainment/affection 
purposes and criminal purposes (for the details see \cite{article}). The
`gods descending from the sky'  could simply correspond to various 
scientific teams, assisted by military personnel, sent by the government
in order to help develop the terrestrial primitive civilization.

\section{ Brane World Cosmologies}\lvm  

To finish I would like to make some comments about brane world cosmologies. 
In the last seven years brane world models have been of increasing interest for
both Particle Physics and Cosmology  \cite{B-W} \cite{RanSun}. They put 
forward the possibility that our Universe with three space dimensions is 
located in a subspace (brane) of a higher dimensional Cosmos. These models 
allow large, and even infinite, extra dimensions\footnote{Previously, in theories 
and models of Particle Physics extra dimensions were only allowed if they were 
compactified with a very tiny radius.} and they offer the possibility to solve, or 
view from a newly different perspective, many longstanding problems in Particle 
Physics and Cosmology (see \cite{reviews} for brane world reviews). 

Brane world cosmologies have the potential to dramatically strengthen the 
Fermi Paradox. Namely, if our observable Universe is embedded in a 
much larger Cosmos there may exist other universes along the extra spatial 
dimensions  which could be parallel to our own, or intersecting it somewhere. 
Then it would be natural to expect that some of these universes would have 
the same laws of Physics as ours and many of the corresponding advanced 
civilizations could master techniques to travel or `jump' through the extra 
dimensions for visitation or colonization purposes. Moreover, one has to take
into account that many of these universes could be very close to ours,
even at only one millimeter distance along an extra dimension.  

This opens up enormous possibilities 
regarding the expansion of advanced civilizations simultaneously through 
several universes with the same laws of Physics, resulting in 
multidimensional empires. It could even happen that the expansion to 
other `parallel' galaxies through extra dimensions could be easier, 
with lower cost, than the expansion inside one's own galaxy. In particular,
the `owners' of the Solar System (if they exist) could
have come from another universe and could have created a huge empire
with large pieces of territory in several `parallel' galaxies.

At present we physicists are still in a premature phase in the study of brane 
worlds and we do not know whether these ideas are in fact realistic. 
Nevertheless, the idea of large extra dimensions and parallel universes is 
acquiring greater and greater importance in the scientific community, among both 
theoreticians and experimentalists. As a matter of fact, the experimental 
signatures expected from large extra dimensions, at present and future colliders, 
are well understood by now \cite{colliders} and an intense experimental search 
is currently under way. For example, experiments starting in 2007 at the LHC 
(Large Hadron Collider) at CERN will be looking, among other things, for 
signatures of large extra dimensions.

\section{Final Remarks}

One final remark. I am convinced that the main feature of the scenario that 
I propose: the existence of primitive subcivilizations embedded in large 
hypercivilizations spanning vast regions of the galaxies, is true at present 
and/or has been true in the past and/or will be true in the future in most or all 
galaxies in our universe and in any other possible universes 
with the same laws of Physics as ours. Whether our terrestrial civilization
is actually a subcivilization completely unaware of the existence of the large  
hypercivilization is something to be discovered in the future. This will happen
either due to some advanced technology that would allow us to discard this 
possibility some hundreds of years from now (if we do not annihilate ourselves
first), or through the decision of our hosts, provided they exist, of showing 
their faces openly, which could happen long before.

\vskip 1.2cm   

\noi
{\large \bf Acknowledgements}

\vskip .3in 
I would like to thank the organizers of the World Mystery Forum 2005 for
inviting me to this beautiful and interesting place. I am also indebted
to my friends Linda Champion and Maite Fern\'andez for many interesting
conversations and also to Linda for the English corrections to this text.

\vskip 1cm   

\noi
{\large \bf Appendix: Questions and Answers}

\vskip 1cm 

\begin{itemize}

\item 
{\bf Why do you say that the open contact with the hypercivilization would destroy 
our civilization?}

This is very easy to understand. Imagine how we would feel if the citizens 
of the hypercivilization happen to live 1.000 years and, in addition,
always looking young. We would feel devastated, right? I can imagine
pathetic demonstrations all over the world demanding `our right' to live 1.000
years as well. What about if they explain to us that we are allowed to live
on planet Earth because of their kindness and hospitality? Also, 
and this is very important, we need
to believe in ourselves, in our capabilities to make progress, to build the
future. This psychological need would crash in contact with a far superior 
civilization and far superior intelligent beings and we would only feel
worthless and stupid, especially scientists, but everybody else as well
(for example, the gap between them and us, regarding brain capabilities, could 
be bigger than the corresponding gap between us and the gorillas). 
Not to mention the shock we would receive if they had created our species
by genetic engineering and they would tell us. It is difficult to predict who
would feel more offended under such circumstances: the fundamentalist
creationists, the intelligent design advocates, the plain hard-core darwinists 
or the ecologists. 

As for myself, I find this last issue quite harmless in fact because, 
as I said, I am completely convinced that advanced civilizations widely use
genetic manipulation on all species around, including themselves. Thus,
if an advanced civilization has created the homo sapiens sapiens, then
I am pretty sure that the individuals of this civilization have also resulted
from genetic manipulations of their own ancestors who in turn have a
high chance to have resulted from genetic manipulation by a much older
alien civilization, who in turn were manipulated by their own ancestors, etc.
It could even happen that at present there is not a single species that remains
`virgin', free of genetic manipulation, in our whole galaxy nor in most 
typical galaxies\footnote{This scenario would constitute the ultimate 
nightmare for ecologists!}. So, regarding the 
possibility that we have been created by scientists of an advanced civilization 
I feel like in that joke where somebody is on the phone: `This is a recording. 
Doesn't bother me, I'm a hologram'.

\item 
{\bf Do you think could it be possible that some of the parallel universes 
that you mentioned could be inhabitated by angels and other spiritual beings?}

I must say that I have been asked this question many times already. At the
present stage, our scientific knowledge can neither prove nor disprove the 
existence of angels and other spiritual beings described in our traditions. 
My opinion is that, if these beings happen to exist, I would expect to find them 
inhabitating parallel universes with {\it different laws of Physics} than ours. 
The crucial point is the laws of Physics that hold in a particular universe. 
Unfortunately we only know one universe and one set of laws of Physics, 
which renders our knowledge very limited and `provincial'. Nevertheless, 
we particle physicists assume that if something would enter our universe 
from another universe with different laws of Physics then the matter of 
this something and the matter of our universe would almost not interact 
(perhaps only gravitationally). As a consequence this something would be 
invisible for most of us and would go through our `solid' matter (walls, ceilings) 
effortlessly, as the angels and other spiritual beings are supposed to do, right?
I say `most of us' instead of `all of us' because the existence of special psychic 
capabilities, such as telepathy, has not been discarded yet and therefore it 
could happen that specially-gifted people could see and/or communicate 
telepathically with beings coming from such universes with different laws of
Physics than ours. These beings would include, in particular, plain
extraterrestrials  (not especially spiritual) coming from other universes just 
for scientific or other mundane purposes. 

As a matter of fact, a good question for a psychically-gifted person (or for a 
cat!)\footnote{Cats have the reputation of having great psychic powers and of
being able to see all kinds of entities invisible to the human eye. 
The aforementioned T. Lobsang Rampa even claimed that his book
 `Living with the Lama' had been written by one of his cats who dictated
 the contents to him telepathically, the introduction being written by
another cat.} would be how he/she/it can distinguish
`who is who' among the entities he/she/it is supposed to see.
I already asked this question to a reader of my article who told me that 
other dimensions and parallel universes truly exist because he can see,
and sometimes communicate with, intelligent beings coming from them.
So I asked him what these beings look like and how he can distinguish among
them. He said: `I see different types of entities, including angels, although 
most of them are human beings in their `astral bodies' (either alive or dead). I
have rarely seen extraterrestrials, but I don't know if they were in their physical 
bodies from another universe, as you suggest. I have the impression that they 
were in their astral bodies, in the same way that astral travellers among us
visit other planets as well.\footnote{The astral 
body is supposed to be able to travel many orders
of magnitude faster than light. This would not contradict any known laws of 
Physics, curiously, because the astral body would not be made out of matter/energy 
of this universe, but out of matter/energy of the `astral universes', with
different laws of Physics. The high speed and low cost of astral travel 
would not be the only advantages with respect to exploration of outer space:
astral astronauts could also approach stars and black holes as much as they 
wanted and even jump inside, returning safely, since the astral body, endowed with 
the senses of sight and hearing with respect to this universe, would neither feel 
gravity nor heat.} One distinguishes these entities 
by their general appearance: body, face, clothing and also the aura. In
addition, alive beings in their astral bodies have some sort of `silvery cord' 
that connects them with their physical bodies, although it can be very
faint and difficult to spot'. 
Then I asked him whether it would be possible to confuse an angel or any other 
`spiritual being' with an extraterrestrial visiting us from this or another universe.
He said: `Yes, it could be possible because angels and all spiritual beings 
create their appearance at will with their minds (face, body, clothing, wings on 
and off, etc.). Angels like to look like handsome human-like beings (whether
terrestrial or extraterrestrial). For this reason angels watching over ugly 
extraterrestrial humanoids also look very ugly, according to our taste, although 
very beautiful for them. So, it would be possible to confuse `ugly' angels with 
equally ugly aliens, and the other way around, very handsome 
extraterrestrials may look like beautiful angels for us'.

\vskip .6cm

 \item 
{\bf There are many reports of `extraterrestrials' who look like pure-energy 
beings of light without bodies. Do you think that all the extraterrestrials could 
be in fact disembodied beings? Also, do you think that in 
the future we could become like these beings of light, too?}

These questions are related to the previous one. First of all, there are also
many reports about extraterrestrials with physical, solid bodies. For example,
today Ed Mitchell, lunar module pilot of Apollo 14, reported that the Roswell
incident was true and one of the aliens on board was still alive when the
spacecraft was
found by the USA military\footnote{In July 1947 the military base near Roswell
(New Mexico, USA) issued a press release announcing that they had
captured an alien spacecraft that had crashed nearby. Three days later they
retracted their statement: the alien spacecraft was only an atmospheric baloon.}.
So I don't see
any reason to think that all possible extraterrestrials around (if any!) must lack
a physical body. On the other hand, I do not believe in the existence of
disembodied beings at all, although perhaps there are beings which are 
visiting our universe from another universe with different laws of Physics. 
This would explain why we could not see their bodies and why they could go
through walls and ceilings! And the other way around, these beings would not
be capable of seeing our bodies either unless they had that psychic capability, 
provided this capability exists at all. So, to me this is very simple: all 
intelligent beings who happen to exist (be human-like, or 
animal-like or angel-like, or whatever-like) must
have a body (at least one!)\footnote{The metaphysical `standard model' says
that human beings (terrestrial and extraterrestrial) have seven bodies: physical, 
etheric, astral, mental and the three superior bodies, each belonging to a 
different universe or `plane of existence'.  Animals would have the four inferior
bodies whereas angels and all other `spiritually advanced' 
beings would have only the three superior bodies.}, no 
matter from which universe they come, and these bodies are solid 
and well visible to the inhabitants of their own universes, 
or another universes with the same laws of Physics as their own. 

About the second question, if you ask if our biological species could evolve
towards `pure-energy beings of light' in this physical universe, the answer 
is no. I do not believe that the laws of Physics of this universe will 
ever allow the emergence of disembodied intelligent beings, as I just said. 

As a matter of fact, your question fits better in metaphysics than in ordinary
science. In the metaphysical `standard model' all conscious beings
(human beings, animals, angels, etc.) are just visitors of the universes they
inhabitate. The reason would be that consciousness cannot be created in any 
universe, no matter the laws of Physics. For example, this implies that nobody 
in any universe could ever construct a 
robot or computer that could become conscious. 
Individual consciousness would have been separated, not created, from
an enormous ocean of consciousness that has always existed, without beginning
and without end, and is the ultimate cause for the existence of everything else.
So, following these views, since our consciousness cannot emerge from any
complex system (such as our brain), it turns out that
we are all aliens in the universes we inhabit and our 
bodies are nothing but the astronaut suits necessary to live in these universes. 
Coming back to your question, in the metaphysical standard model, we human 
beings are supposed to live many lives in `physical universes', where the mind
has very little power over matter, working very hard, and resting 
in between in `astral universes', where it is always vacation time 
because the mind can create all the basic needs: housing, furniture, clothing,...
Now, after many incarnations in the physical and astral universes we can be
promoted to better, more spiritual universes where the power of the mind over
matter is much stronger than in the astral universes. This way, we would
become spiritual `beings of light' without any physical, etheric and astral 
bodies, although we would still have the more superior bodies made out of the 
matter/energy of the more spiritual universes we would inhabit. 
These bodies would be invisible for most inhabitants of the physical 
universes and also for most inhabitants of the astral universes.    

\vskip .6cm

\item 
{\bf Did you have any unusual experiences yourself that motivated the writing
of your article?} 

Not directly; I have never seen a UFO, nor have I tried to see one or to get in 
contact with aliens, as some people do. Indirectly, I had a most unusual experience
more than twenty years ago that has been a source of inspiration indeed, although
the true motivation to write the article was simply to reply to the ideas of the
cosmologist Ken Olum, as I explain in detail in the article. 

\vskip .6cm

\item 
{\bf What kind of experience did you have? Could you say something about it?}

This is a rather long story.  In May 1984 a cousin of mine invited me for
coffee at her place and she also invited an old friend of hers. She warned 
me about the possibility of hearing strange statements because her friend was 
in contact with extraterrestrials for several years already. At her place the
conversation was revolving around ordinary matters. He was a completely 
normal-looking guy with a normal job. After one hour or so we brought up the
subject of summer vacations. At his turn he said, with a completely normal 
voice and straight face: `I will go with my friends in July, we have already 
arranged things. We will be a couple of days in the bases of the Moon 
and then they will bring me three weeks to 
Ganymede\footnote{Ganymede, one of the
moons of Jupiter, is also the largest satellite in the Solar System, larger
than the planets Mercury and Pluto.} again. Last summer I spent some days in
Confraternity City, where there is a terrestrial colony of about 12.000 people.
They have a very interesting museum of History of Earth, where one can see
the bodies of the prophet Elijah and the patriarch Enoch. They lived in 
Ganymede for hundreds of years until they died'.

Although I didn't believe a word and I really thought he was crazy, I tried to
show some interest, mainly out of politeness (he was very polite himself), 
and so I started asking some
questions. I asked how the view of Jupiter was from Ganymede, how long it
takes to travel to the Moon and to Ganymede, how many people inhabit the
satellite and what his friends look like. His answers were: `Jupiter is enormous,
it covers almost the whole sky', `The trip to the Moon takes only two or three
minutes, to Ganymede it depends on the spacecraft. In standard ones the
trip takes a bit more than three days whereas in cylinder-shaped motherships
the trip takes only three hours'. `There are about two and a half million people
in Ganymede, that they call Morlen, distributed mainly in five cities. In Crystal
City is the government of Morlen as well as the government of the Confederation
to which Morlen belongs. It is a Confederation of the 24 more advanced worlds
in our galaxy and the government is called the Council of the 24 Elderly'.
`My friends from the bases on Earth are mainly from Ganymede, although 
one can meet people from the other 23 worlds as well. Most
of them look similar, although taller or shorter depending on which world they
come from. In general they are very handsome (angel-looking faces) with 
long blond hair, white skin and light-colored big eyes (a bit oblique), 
and most of them look like
40 years old, although they can be 300, 600, 800 years old since they can 
live until 1.200 (terrestrial) years and they master anti-aging technology. 
The most impressive-looking ones
are the giants from planet Apu, in Alpha Centauri, who are about three meters
tall and in addition wear very long, almost white hair. My friends from Ganymede 
are also quite tall: more than two meters'.

At that moment I replied back: `So, they are quite similar to us, although taller and
more handsome, and in addition they are the most evolved people in our galaxy.
What a coincidence! How lucky we are that we have such wonderful neighbors!'
Then he said with a serious expression: `This is not a coincidence, I am affraid,
but I prefer not to give you the details. You are still too young and you 
wouldn't be able to handle the facts. Just let me tell you that they arrived at this 
part of the galaxy: Alpha Centauri, the Solar System, the Pleyades, etc. almost 
three million years ago. They established themselves in artificial colonies without
any natural life, so they constructed their worlds completely from scratch. They
didn't bring animals along and therefore they only have plants and are vegetarian'.

Then I asked him how he got in contact with his friends and how they could 
understand each other. He said: `They contacted me; I had never been 
interested in UFO's and extraterrestrials'. `They have three glands inside the
brain that we don't have. One of them produces very strong telepathic
capabilities, so they do not speak with the mouth, only by telepathy. It
is very easy to understand them because they speak a `universal language'
that is converted inside the receptor's brain into his/her mother tongue. 
For this reason they don't use technological devices for communications 
either, like phones or radios, they communicate
exclusively by telepathy, even from one planet to another'.

Then he gave us the advice that we should never try to get
in touch with aliens: `There are some very dangerous humanoids, who come from
another universe, that do horrible things to us. Although some groups of people
have been lucky and have entered into telepathic communication with people from 
the Confederation, including some of my friends, the risk of getting the wrong 
extraterrestrials is high'. Then I asked:
`What do you mean by another universe, do you mean that they come from another
galaxy? No, I mean another universe. There are 22 dimensions, not just the three
that we know.  For this reason there are lots of universes. These guys live in 
a universe nearby which has an entrance to our universe very close to the Solar 
System. Once they enter they only have to travel for three days to reach Earth. 
They are not allowed to come here, of course, but they do it anyway.  So, these 
guys are a real nightmare for my friends who do their best to chase them away. 
In fact, they have already crashed more than once during the persecutions because
their technology is far below the technology of the Confederation. Otherwise
they would have taken over the planet and made us their slaves'. 
Finally I asked him whether all the universes are like ours.
He said: `The universe of the intruders is like ours, unfortunately, but
as far as I have heard, many universes are different'.

I never saw him again (nor did I want to!). I had no doubts that he had a deep
psychiatric illness whereas my cousin defended him saying that she had known
him very well for many years and he was neither a liar nor crazy, therefore the
story had to be true, no matter what. 

For the following fourteen years or so I did not think at all about the subject of 
extraterrestrial civilizations. Unexpectedly, one day around 1998 the 
subject came up somehow when talking to my friend Maite at her place. 
Then she showed me a couple of curious books about alleged contacts. 
One of the books, `Los Manuscritos de Geenom (II)' (The Manuscripts of
Geenom (II)) was written by a group based in Madrid, the Aztl\'an 
group, which claimed to be in telepathic communication, once per week 
and for about twenty years already, with some citizens of planet 
Apu\footnote{In the `communications', which started in 1975 and still continue, 
it is reported that Apu orbits the star Alpha B Centauri, at a distance of 
4.35 light-years from the Sun. Alpha Centauri, which is the closest stellar 
system to us, consists of a binary system of stars with sizes similar to
that of the Sun - Alpha A (yellow) and Alpha B (orange) - accompanied at a 
much greater distance by a dim red dwarf, Proxima Centauri, which is the 
nearest star to the Sun in our observable universe, 4.22 light-years away.
One can find interesting astronomical information about Alpha Centauri in the
web site http://homepage.sunrise.ch/homepage/schatzer/Alpha-Centauri.html} 
(the `communications' were supposed to be transmitted
from planet to planet). The other book had for its title 
`Yo visit\'e Gan\'\i medes' (I visited Ganymede), by Yosip Ibrahim. 
So, suddenly I got `teletransported' back to the coffee session fourteen years 
earlier. I read these two books and, for the first time, I realized that I had 
never spent a single minute thinking seriously about the subject of
extraterrestrial civilizations, even though I had always been open-minded 
regarding the possibility of alien visitors. I became aware that this issue
could be much more important and have much more far-reaching consequences 
than I, and most open-minded people, naively could have imagined.  
   
After spending some time thinking about this issue I reached the conclusion that
it was not possible to discard the scenario described by the friend of my cousin,
no matter how strange it seemed to me. There were only two crucial questions 
to be answered:  Would a civilization, millions of years ahead of us, be 
able to colonize places like the large satellites of Jupiter? and, would such
a civilization be able to completely conceal itself from external observers?
My answer to these two questions was undoubtedly positive.

For the following year Maite and I read a few more books \cite{books} 
about the Confederation-Apu-Ganymede affair, for our own records, 
and that was all. It never crossed my mind to write an article about primitive 
versus advanced civilizations, and even less to try to find out {\it the truth} 
about the whole issue\footnote{This is left for the interested reader.}. 
However, in March 2003 the article by Ken Olum \cite{Olum} 
appeared and triggered my attention again to the 
subject. He was saying that our small terrestrial civilization should be part of 
a galactic civilization spanning a large region of the galaxy, as followed from 
his computations, but however we are not part of such a civilization, as 
observation confirms. Needless to say, this article stirred my mind notably,
even though I didn't agree with Olum's arguments and computations. So I 
started thinking seriously again about this issue of extraterrestrial 
civilizations and shortly afterwards I had the ideas of the Subanthropic 
Principle and the Undetectability Conjecture.  Then I decided to write an 
article discussing the possibility that our terrestrial civilization could be 
embedded in a large advanced civilization without being aware of it.

\end{itemize}

\end{document}